\documentclass[prl,floatfix,superscriptaddress,twocolumn,aps,showpacs,amsmath,amssymb]{revtex4}
\usepackage{graphicx}% Include figure files
\newcommand{\e}{\ensuremath{\mathrm{e}}}
\newcommand{\VSD}{\ensuremath{V_{\textsc{sd}}}}
\newcommand{\VSDbar}{\ensuremath{\bar{V}_{\textsc{sd}}}}
\newcommand{\ColorOnline}{(Color online) }
\newcommand{\be}{\begin{equation}}
\newcommand{\ba}{\begin{eqnarray}}
\newcommand{\ea}{\end{eqnarray}}
\newcommand{\ee}{\end{equation}}

\newcommand{\Cal}[1]{{\cal #1}}
\newcommand{\p}{\partial}

\newcommand{\XXX}[1]{}

\newcommand{\bea}{\begin{eqnarray}}
\newcommand{\eea}{\end{eqnarray}}

\newcommand{\f}{{\tiny \mbox{F}}}

\newcommand{\Marb}{m}
\newcommand{\Ucont}{U_c}
\newcommand{\tcont}{t'_c}

\begin{document}
\title{Twofold advance in the theoretical understanding of far-from-equilibrium properties of interacting nanostructures}
%Out of equilibrium transport in interacting nanostructures: a double step forward
\author{E. Boulat}
\affiliation{Laboratoire MPQ, 
CNRS UMR 7162, Universit\'e Paris Diderot, 75205 Paris Cedex 13}
\author{H. Saleur}
\affiliation{Institut de Physique Th\'eorique, CEA, IPhT and CNRS, URA2306, Gif Sur Yvette, F-91191}
\affiliation{Department of Physics, University of Southern California, Los Angeles, CA 90089-0484}
\author{P. Schmitteckert}
\affiliation{Institut f\"ur Nanotechnologie, Forschungszentrum Karlsruhe, 76021 Karlsruhe, Germany}

\begin{abstract}
We calculate the full $I\!\!-\!\!V$ characteristics at 
vanishing 
temperature in the 
self-dual interacting resonant level model
 in two ways. The first uses careful time dependent DMRG with large number of states per block and a representation of the reservoirs as leads subjected to a chemical 
potential. The other
%potential, the other
is based on integrability in the continuum limit, and generalizes early work of Fendley Ludwig Saleur on the boundary sine-Gordon model. The two approaches are in excellent agreement, and uncover among other things a power law decay of the current at large voltages when $U>0$.

\end{abstract}
\maketitle
%%
%%
%%%%%%%%%%%%%%%%%%%%%%%%%%%%%%%%%%%%%%%%%%%%%%%%%%%%%%
%% DMRG draft
%%%%%%%%%%%%%%%%%%%%%%%%%%%%%%%%%%%%%%%%%%%%%%%%%%%%%%
%%

Experimental investigation of transport phenomena in quantum impurities is a rapidly expanding field. Typically, tiny structures that behave quantum mechanically -- i.e.~, whose level spacing is much larger than all other relevant energy scales -- are connected to metallic leads and a voltage $\VSD$ across it forces a current to flow. The non equilibrium regime that ensues could be achieved and measured for structures realized as quantum dots in a 2D electron gas \cite{revQD} or single molecules \cite{revMol}.

Unfortunately, the theoretical description of these 
%situations 
systems
remains, in spite of numerous efforts, somewhat less advanced. One major obstacle is the lack of efficient 
theoretical 
approaches to treat non-equilibrium situations in the presence of strong interactions. On the analytical side, perturbative
(Keldysh) techniques are extremely difficult to carry out to high orders, while self-consistent approximations are
difficult to control in the strongly non linear regimes.
Since impurity problems in the scaling limit can be reformulated as 
%one dimensional 
1D
boundary field theories, it is tantalizing to try to use 
%here 
the power of integrability. This was done first in \cite{fendley95a} where full $I\!\!-\!\!V$ characteristics were calculated for the problem of edge-state tunneling in the fractional quantum Hall effect. More recently, Mehta and Andrei \cite{mehta06} have proposed a seemingly different approach dubbed the open Bethe ansatz. They also cast doubt about the results in \cite{fendley95a} and their way of coupling the model to the reservoirs. While the approach in \cite{mehta06} is a priori quite general, it has not, so far, led to quantitative  predictions
 in  the scaling limit because of technical difficulties. 
 %Moreover, 
 Besides,
 since the cut-off in \cite{mehta06} is imposed via the Bethe ansatz solution, the results cannot  be directly  compared with those of lattice simulations.

In view of this confusing situation, it would be natural to turn to numerical approaches.
This is however just as challenging.
In simulating quantum transport one faces the problem
that the stationary Schr\"odinger equation
is replaced by the time dependent Schr\"odinger equation
and therefore an eigenvalue problem is replaced by a
boundary problem in time. An important consequence is that in simulations
in time (frequency) domain one has to send first system size to infinity and then
time to infinity (level broadening/frequency to zero).
Current numerics typically falls into two classes.
Either one takes the limit of switching on interaction last and starts from a noninteracting description,
where one can send the system size analytically to infinity. Or one is using nonpertubative
methods on a finite lattice where the limit of time/frequency is interchanged with the limit of system size.
Therefore numerical simulations for transport properties of strongly interacting quantum system are in
general prone to conceptual considerations.

We report in this letter a double step forward. We on the one hand extend the approach of \cite{fendley95a} to a special point in the interacting resonant level model (IRLM), where we are able to determine the full $I\!\!-\!\!V$ characteristics at vanishing temperature $T$. We on the other hand carry out time dependent DMRG (td-DMRG) calculations, and fully confirm our Bethe ansatz predictions in the scaling limit. The remarkable agreement between analytical and numerical approaches strongly validates both, hence dispelling doubts about the issues of reservoir coupling in \cite{fendley95a} as well as the feasibility of time dependent DMRG. On top of this, the physics unraveled by our calculations is highly non perturbative and counter intuitive, and exhibits, among other things, regimes of negative differential conductance. We thus expect the IRLM model to become a benchmark for other methods in the field of interacting out of equilibrium transport.

We start with the numerical approach.
 We apply the td-DMRG method\cite{White:1992,DMRG:Reviews,Schmitteckert:2004,White_Feiguin:2004,Daley:2004} to
integrate an initial state of the nanostructure ${\cal H}_{\mathrm I}$
attached to a left and a right noninteracting tight-binding lead ${\cal H}_{\mathrm L}$
\bea
{\cal H}_{\mathrm L} &=& -t
\big(\textstyle{\sum_{x=2}^{x_0-1} + \sum_{x=x_0+2}^{M}} \big) 
(\hat{c}^\dagger_x \hat{c}^{}_{x-1} \,+\, \hat{c}^\dagger_{x-1} \hat{c}^{}_{x})
\vphantom{\bigg(}
\\
 {\cal H}_{\mathrm I} &=& t'\, \hat{c}^{\dagger}_{x_0} \left( \hat{c}^{}_{x_0-1} \,+\,\hat{c}^{}_{x_0+1} \right) \;+\; \text{h.c.}\label{impinter}
\\
&+& U\, (\hat n_{x_0} - 1/2)*(\hat n_{x_0-1} + \hat n_{x_0+1} -1) \,+\, \epsilon_d \hat n_{x_0}\,,
\nonumber
\eea
where $t=1$ is the hopping
amplitude 
in the leads, $M$  the number of lattice sites,
$U$ the interaction on the contact link and $\epsilon_d$  an on-site gate potential which
is set to zero (i.e. at resonance).

At time $\tau=0$ we include a voltage drop by applying a potential \VSD/2 (-\VSD/2) on the left (right)
lead
which we smoothly send to zero on a scale of three sites left and right the impurity.
We then switch off the 
voltage
in the Hamiltonian and 
time evolve
 using
$\e^{ i \hbar ( {\cal H}_{\mathrm L} + {\cal H}_{\mathrm I})\tau}$ 
(different procedures to reach the stationary state could be considered. We will report on this elsewhere - see also \cite{DoyonAndrei}).
Since the leads act as a bath the simulation
has to be stopped after the transit time $\tau_{\rm t}=L_{\text{lead}}/v_\f$, where
 $L_{\text{lead}}$ is the leads' length and $v_\f=2t$ is the Fermi velocity
of the noninteracting tight binding leads, since for times larger than $\tau_{\rm t}$
one measures the influence of the boundaries and not the steady state.
In addition, since we are working with finite leads, our system has a finite size
gap which leads to finite size induced $\cos( \VSD\, \tau)$ oscillations.
This is similar to the
oscillations in a Josphson junction which are induced by the superconducting gap.
By carefully checking for finite size effects and ensuring to take enough states
per block in the DMRG procedure we can extract the current corresponding to infinite leads,
the details are explained in \cite{Schmitteckert:2006,Schmitteckert:2007}.

\begin{figure}[ht]
\begin{center}
    \includegraphics[width=0.475\textwidth]{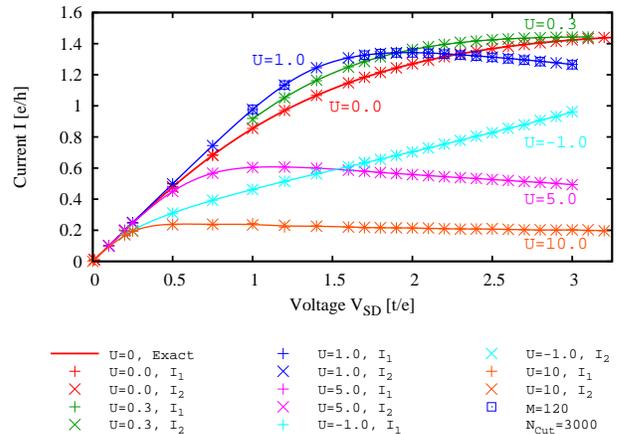}
    \caption{\ColorOnline Current vs. source drain voltage \VSD\ for
    a hybridization of $t'=0.5$, and interaction values of $U=-1.0, 0.0, 0.3, 1.0, 5.0,$ and $10.0$.
    The subscript '1' ('2') of $I$ refers to data extrapolated from link to
    the left (right) lead. The lines are guides to the eye, except for $U=0.0$ where we
    plotted the exact result for infinite leads.
    Calculation have been performed with 96 sites, 48 fermions keeping at least 2000 states
    per DMRG block. In addition data for $U=1.0$, $M=120$ sites and $N_\text{Cut}=3000$ states per block is shown.
 }
   \label{IvsV}
\end{center}
\end{figure}

In Figure~\ref{IvsV} we plot the current $I$ vs.\ applied source drain voltage
for a system with $t'=0.5$ and $\epsilon_d=0$, i.e.\ on resonance.
Most of the data was calculated using a 96 site system
and at least 2000 states per block.
For comparison we include a reference calculation for $U=1.0$ using 120 lattice sites
and $3000$ states per block. We applied typically 20 to 25 full td-DMRG steps with a time step
of $\Delta_\tau=0.4$ as described in \cite{Schmitteckert:2004}. We then switch to an adaptive
time evolution scheme as described in \cite{DMRG:Reviews,Schmitteckert:2007}. We would like to remark
that we perform the time evolution using a full Arnoldi type matrix exponential during
the full and adaptive td-DMRG sweeps without any Trotter like approximation schemes.

Since free fermions provide a non-trivial test for (real space) DMRG the $U=0.0$ results
show that the procedure is well defined and gives accurate results even for large voltages.
Switching on the interaction one observes in the not too large \VSD\ regime
a broadening of the differential conductance while for large interaction the broadening is absent
and one obtains a shrinking of the resonance width. This is similar to the linear conductance
vs.\ gate potential as described in \cite{Bohr_Schmitteckert:2007}.
For large voltage a negative differential conductance regime appears which is maximal at $U\sim 2.0$
and disappears again for $U\rightarrow\infty$.
We note that DMRG has a tendency to underestimate the current in the
large voltage regime if not enough states are kept, especially in the adaptive scheme.

\begin{figure}[ht]
\begin{center}
    \includegraphics[width=0.475\textwidth]{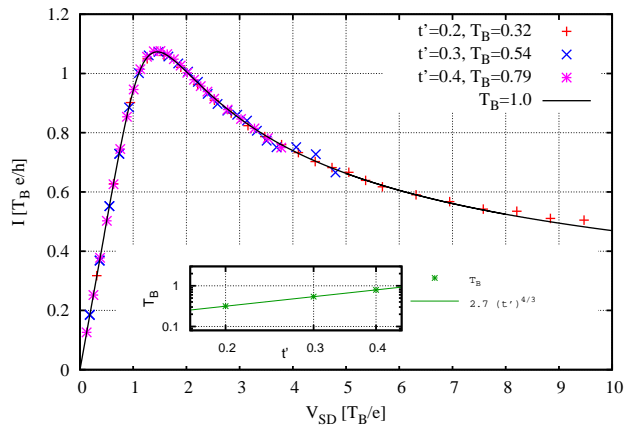}
    \caption{\ColorOnline
Comparison between analytical and DMRG results at the self dual point. The numerical data
have been fitted using a single parameter $T_B = c (t')^{4/3}$, $c \approx 2.7$.}
   \label{Universal}
\end{center}
\end{figure}

In Figure~\ref{Universal} we plot the IV curves for $t'=0.2,0.3,0.4$ where we rescaled
the $I$ and $\VSD$ axis in such a way that we fit the data to our analytical result using a scale $T_B$.

The points nicely sit on a single curve, which shows that we are reasonably within the scaling limit.
A numerical fit of the power law decay for $t'=0.2$ and $U=2.0$ gives an exponent of 0.47.
In the inset of Figure~\ref{Universal} we show that the numerically obtained $T_B$ scales
are given by a $T_B \sim (t')^{4/3}$ power law.

\begin{figure}[ht]
\begin{center}
    \includegraphics[width=0.8\linewidth]{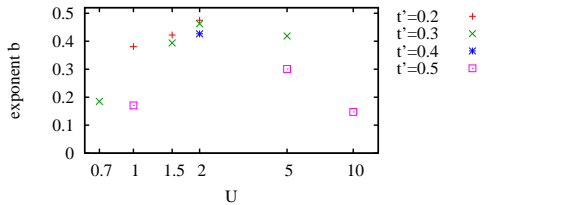}
    \caption{\ColorOnline Exponents in the negative differential conductance regime for $t'=0.2, 0.3, 0.4$, and $0.5$.
    }
   \label{Exponents}
\end{center}
\end{figure}

Exponents for different values of $U$ are represented on figure~\ref{Exponents}.
For all positive values of $U$ we find a power law decay of the current at large voltage,
except for $t'=0.5$ and $U=0.3$ where the crossover scale is beyond our voltage regime.
We also find that the exponent does not vary monotonously with $U$, and reaches its maximum around $U=2$ where it is very close to $1/2$, and approaches $1/2$ when lowering $t'$, i.e. in the scaling limit.

We now turn to a study of the model in the continuum limit.
It is convenient to first unfold the left and right leads, obtaining in this way ${\cal H}_{\mathrm L}= -i\sum_{a=1,2}\int dx\, \psi^\dagger_a\p_x\psi^{}_a$
a free Hamiltonian describing
two infinite right moving Fermi wires, and
\begin{eqnarray}
& {\cal H}_{\mathrm I} = \tcont\left(\psi^{}_1(0)+\psi^{}_2(0)\right)\,d^\dagger +
\mbox{h.c.}
\vspace{0.1cm}\label{hamiltonian}
&\\
%H_B = \tcont\left(\cos\frac{\theta}{2}\psi^{}_1(0)+\sin\frac{\theta}{2}\psi^{}_2(0)\right)\,d^\dagger +
%\mbox{h.c.}
%\vspace{0.1cm}\\
& +\;\Ucont \;
\big(:\psi^\dagger_1\psi^{}_1(0):+:\psi^\dagger_2\psi^{}_2(0):\big) \, \big(d^\dagger d-\textstyle\frac{1}{2}\big) + \epsilon_{d}~d^\dagger d 
& \nonumber
\end{eqnarray}
In the following, we focus again on the resonant case, i.e.~$\epsilon_d=0$. Note that in (\ref{hamiltonian}) the Coulomb interaction involves normal ordered charge density on the dot, as in (\ref{impinter}). The subscript $"c"$ indicates that the precise relationship between the lattice and continuum limit parameters depends on the regularization scheme.

We switch to the language of the anisotropic Kondo model by representing the impurity degree of freedom by a spin one half, $d^\dagger=\eta S^+$, $S^z=d^\dagger d-\frac{1}{2}$, with $\eta$ a Majorana fermion. The fermions in the wires are bosonized as $\psi_a=\frac{\eta_a}{\sqrt{2\pi}}\,e^{i\sqrt{4\pi}\varphi_a}$. We then introduce new bosonic fields $\varphi_c=(\varphi_1+\varphi_2)/\sqrt{2}$ and $\phi=(\varphi_1-\varphi_2)/\sqrt{2}$, that represent the total and relative charge in the wires respectively.
The unitary transform $\Cal U=\exp\big(i\alpha\sqrt{2}\,S^z\,\varphi_c(0)\big)$ is then applied. Choosing $\alpha=\Ucont/\sqrt{\pi}$ to cancel the boundary interaction along $S^z$, we arrive at:
\be
{\cal H} = {\cal H}_0(\varphi_c) + {\cal H}_0(\phi) + {\cal H}_{\mathrm I}
\label{hamcont}
\ee
where ${\cal H}_0(X)\!=\!\int_{-\infty}^\infty dx\,(\p_x X)^2$
is the Hamiltonian of a free (chiral) boson,
and the piece involving the impurity degrees of freedom is ${\cal H}_{\mathrm I}=
\frac{\tcont }{\sqrt{2\pi}}\,\left[\Cal V(0)\Cal O(0)
S^+ + \mbox{h.c.} \right]$, with
$\Cal V\!=\!e^{i\beta_c\varphi_c}$,
$\Cal O\!=\!\kappa_1 e^{i\sqrt{2\pi}\phi} \!+\! \kappa_2 e^{-i\sqrt{2\pi}\phi}$,
and $\kappa_a\!=\!\eta\eta_a$. The exponent of the vertex operator is $\beta_c=\sqrt{2\pi}(1-\frac{\Ucont}{\pi})$. We note that there is a special value of the coulombic repulsion 
($\Ucont\!=\!\pi$
in our scheme) where a remarkable simplification occurs: this exponent vanishes, and the field $\varphi_c$ decouples from the impurity. At this point, and this provides a universal characterization 
thereof, the anomalous dimension of 
the boundary perturbation 
$D=\frac{1}{4} \!+\! \left(\frac{1}{2}-\frac{\Ucont}{2\pi}\right)^2$ 
reaches its minimum value $D\!=\!\frac{1}{4}$. At this point, the model exhibits a certain self-duality \cite{SchillerAndrei}.

Although we do not have full analytical solutions for general values of $\Ucont$, some qualitative results can easily be obtained. First, we expect that, provided $\Ucont>0$, the current vanishes at large voltages like a power law
$I_0(V)\sim \VSD^{-b}$, $b=1-2D$, $D$ the scaling dimension of the perturbation (this conclusion was reached in discussions with B. Doyon \cite{Benjamin}). From the foregoing discussion this leads to an exponent $
b={1\over 2}{\Ucont\over\pi}\left(2-{\Ucont\over\pi}\right)$
We emphasize that this only holds in our regularization scheme. The numerical values of $\Ucont$ where $b$ reaches its maximum and where $b$ vanishes do not have to be the ones observed in the lattice model; in particular $\Ucont=2\pi$ for us corresponds presumably to $U=\infty$ where it can be argued that the model is equivalent to $U=0$. Nevertheless, it is usually expected that different regularizations do not change the qualitative nature of the results. The prediction for the exponent is indeed in full agreement with numerics and the curve on figure \ref{Exponents}. We see in particular that the exponent $b$
does exhibit maximum value at $b={1\over 2}$ where $D={1\over 4}$ by scaling. This allows us to identify the self dual point in the continuum limit with the value $U\approx 2$ in the lattice model. We shall now see that this self dual point is also amenable by the Bethe ansatz.

The IRLM in equilibrium is solvable:
the corresponding basis of the Hilbert space provides quasiparticle excitations which scatter diagonally (ie, without particle production) across the impurity. In general however,
the operator enforcing the voltage drop across the impurity ${\VSD\over 2}\int dx\big(\psi^\dagger_1\psi_1-\psi^\dagger_2\psi_2\big)\equiv {\VSD\over 2}Q$ is \emph{not} diagonal in the quasiparticle basis. This makes the construction of exact scattering states along the lines of \cite{fendley95a} a seemingly impossible task. At the special value of the coupling $\Ucont\!=\!\pi$ however, things simplify drastically. The hamiltonian can be mapped onto the boundary sine-Gordon theory for the field $\phi$ after folding across the impurity (it can be shown that the cocycles can be discarded at this stage). This theory is known to be integrable, and leads to a description in terms of solitions, antisolitons, and two kinds of breathers. The charge $Q\!=\!\sqrt{\frac{2}{\pi}}\int \p_x\phi$ is not conserved by the interaction, but acts diagonally on the quasiparticle basis. Following the arguments in \cite{fendley95a} this allows for an exact calculation of the non equilibrium current at all values of $T$ and voltage for the IRLM at $\Ucont\!=\!\pi$. We will simply give here the result at $T\!=\!0$.

In this case, the in-state involves one type of quasiparticle - for $\VSD>0$ say it is the antisolitons. We parametrize their energy (equal to the momentum) as $p={\Marb\over 2} e^\lambda$ where $\Marb$ is an arbitrary mass scale, and $\lambda$ the rapidity. The antisoliton filling fraction thus reads $f_-(\lambda)=\Theta(A-\lambda)$ ($\Theta$ is the Heaviside function), $A$ being a rapidity cut-off.
Defining $\rho=n f_-$, where $n$ is the density of allowed states per unit of length and rapidity, the current reads (we use units in which $e=\hbar=1$)
\be
I=2\int_{-\infty}^A d\lambda\, \frac{\rho(\lambda)}{1+e^{6(\lambda-\lambda_B)}}.
\label{I0}
\ee
Here, $\lambda_B$ is a rapidity encoding the crossover energy scale in the problem
 $T_B={\Marb\over 2}e^{\lambda_B}$. From scaling, $T_B\propto (\tcont)^{4/3}$. Like for the Kondo temperature, different ways of defining $T_B$ are possible. We follow here the definition given in \cite{lettlin} which is related to universal coefficients in the 
low-$T$
expansion of the linear conductance; we will not need it in the following anyway. The density $n$ -- or equivalently $\rho$ -- follows from non trivial Bethe ansatz quantization rules which involve the antisoliton-antisoliton scattering matrix. The Wiener Hopf technique allows to obtain the Fourier transform $\tilde\rho(\omega)=\int_{-\infty}^A d\lambda\,e^{i\omega\lambda}\rho(\lambda)$ in closed form.
One finds (see Ref. \cite{fendley95b}):
\be
\tilde \rho(\omega)=\frac{\Marb}{4\pi}
%\textstyle{
\frac{\Cal G(\omega)\Cal G(-i)}{1+i\omega}
%}
e^{(1+i\omega)A}
\ee
with
$\Cal G(\omega)=\sqrt{8\pi}\; \frac{\Gamma\left(\frac{2i\omega}{3}\right)}{\Gamma\left(\frac{i\omega}{6}\right)
\Gamma\left(\frac{1+i\omega}{2}\right)} \; e^{i\omega\Delta}$
 and $e^\Delta=\frac{\sqrt{3}}{4^{2/3}}$.
The relation between the cutoff $A$ and the applied voltage can be expressed through the Fermi momentum $p_\f={\Marb\over 2}e^A=
\frac{2^{1/3}}{3^{1/2}}\frac{\Gamma(1/6)}{\Gamma(2/3)}\;\frac{\VSD}{2\sqrt{\pi}}$.
Power expanding the denominator of the integrand 
in (\ref{I0}) yields an 
explicit series representation with
natural expansion variable
$\VSDbar=\frac{\Gamma(1/6)}{4\sqrt{\pi}\Gamma(2/3)}\,\frac{\VSD}{T_B}$
\cite{fendley95b}. 
Depending on whether $\VSDbar$ 
is small or large 
one has:
\bea
I_0(\VSD) & \displaystyle{ \mathop{=}_{\VSDbar<e^\Delta} } & \VSD \;
%\stackrel[n\geq 0]{}{\textstyle{\sum}}
\sum_{n\geq 0} 
\textstyle{\frac{(-1)^n}{4\sqrt{\pi}}}
\displaystyle{\frac{(4n)!}{n!\Gamma(3(n+\frac{1}{2}))}}
\;\; \VSDbar^{6n}\hspace*{0.6cm}
\label{IVsmall}   
\\
I_0(\VSD) & \displaystyle{ \mathop{=}_{\VSDbar>e^\Delta} } & \VSD
%\stackrel[n> 0]{}{\textstyle{\sum}}
\sum_{n> 0}
\textstyle{\frac{(-1)^{n+1}}{4\sqrt{\pi}}}
\displaystyle{\frac{\Gamma(1+\frac{n}{4})}{n!\Gamma(\frac{3}{2}-\frac{3n}{4})}}
 \;\; \VSDbar^{-\frac{3n}{2}}\hspace*{0.6cm}
\label{IVlarge} 
\eea

For comparison, recall that in the free fermion case ($U$$=$$\Ucont$$=$0) one has
$I_0(\VSD)=\frac{T_B}{\pi}\,\hbox{Atan}\big( {\VSD\over 2T_B}\big)$.
We see in (\ref{IVsmall},\ref{IVlarge}) 
that $I_0/T_B$ is a function of $\VSDbar$ only, so that 
matching the numerical results in the scaling limit
requires a \emph{single} common rescaling of 
$I$ and $\VSD$.
The $T=0$ current is depicted in figure~\ref{Universal} (solid line).
It decays as a power-law at large voltage,
$ I_0 \simeq \frac{3^{3/4}}{8\pi}\frac{\Gamma(2/3)^{9/2}}{\Gamma(3/4)^{2}} \;\;T_B^{3/2}\VSD^{-1/2}$.

We can understand more precisely the origin of the large $\VSD$ decrease of the current by looking at the non-equilibrium density of states for anti-solitons:
$\rho(p)=F_\rho(\frac{p}{p_\f})\Theta(p_\f\!-\!p)$ with
$F_\rho(x)=\frac{3}{8\pi^2}$$\frac{\Gamma(\frac{2}{3})}{\Gamma(\frac{1}{6})}$
$\sum_{n\geq 0}$
$\frac{(-)^n}{(2n+1)!}$
$\Gamma(\frac{2m+5}{4})\Gamma(\frac{6m+1}{4}) (\frac{x}{e^\Delta})^{\frac{6m+1}{2}}$.
At small momentum $p\ll\VSD$
-- where the resonance (of width $T_B$) forms -- the density of states vanishes as a power law, $\rho(p)\sim \sqrt{p/\VSD}$: this depletion of the sea close to zero energy results in an extinction of the current.
The situation is clearly contrasted to what happens in a free theory ($U=0$), where the density of states is \emph{constant} provided $p<p_\f=\VSD$.

We note that the formula $b(\Ucont)$ suggests a power law divergence of the current at large voltage in the scaling regime for $\Ucont\!<\!0$. This seems confirmed by our numerics.

 In conclusion our work provides what may be the first example of
 transport properties in an interacting one dimensional system out of equilibrium that can be calculated both analytically and numerically, with excellent agreement between the two approaches. This should provide a most useful benchmark for the variety of other approaches being currently proposed. Our results exhibit remarkable physics: the negative differential conductance at large voltage seems a truly non perturbative behaviour,
with unclear physical origin:
a possible explanation could be that once we are in the tail of the conductance curve a voltage
  drop at the impurity builds up since differential conductance is now smaller than unity. This voltage drop
  may destroy the interaction based renormalization of the conductance enhancement.

The IRLM provides a perfect laboratory to explore other challenging questions, such as the DC shot noise, or the effect of coupling of the baths on the stationary properties. We hope to get back to these soon.
  
 \noindent{\bf Acknowledgments}: 
we are grateful to
N. Andrei, B. Doyon, F. Essler and P. Mehta for numerous discussions and encouragments. HS was supported by the ESF network INSTANS.
 The DMRG calculations have been performed on HP XC4000 at Steinbuch Center for Computing (SCC) Karlsruhe under project RT-DMRG.

%%
%%%%%%%%%%%%%%%%%%%%%%%%%%%%%%%%%%%%%%%%%%%%%%%%%%%%%%

\end{document}